\documentclass[journal]{IEEEtran}
\usepackage{latexsym}
\usepackage{amsmath}
\usepackage{amssymb}

\def \Z {{\textrm{Z}}}
\def \F {{\textbf{F}}}
\newtheorem{lemma}{Lemma}

\newtheorem{theorem}{Theorem}

\def \epf {\hfill $\Box$\\ \par}

\title{A Construction for Constant-Composition Codes}

\author{Yang Ding \thanks{This work was supported by the China Scholarship Council.
}
\thanks{The author is with the Department of Mathematics, Southeast
University, Nanjing, 210096, People's Republic of China (e-mail:
PG23067461@ntu.edu.sg). This work was carried out while the author
was studying in Division of Mathematical Sciences, School of
Physical and Mathematical Sciences, Nanyang Technological
University, Singapore under the exchange program.}
\thanks{}}

\date{}

\begin{document}

\maketitle


\begin{abstract}By employing the residue polynomials, we give a construction of constant-composition codes. This construction generalizes the one proposed by
Xing \cite{b1}. It turns out that when $d=3$ this construction
gives a lower bound of constant-composition codes improving the
one in \cite{b2} for some case. Moreover, for $d>3$, we give a
lower bound on maximal size of constant-composition codes. In
particular, our bound for $d=5$ gives the best possible size of
constant-composition codes up to magnitude.

\end{abstract}


\section{INTRODUCTION}

Constant-composition codes are a subclass of constant weight
codes, in which both weight restrict and element composition
restrict are involved. The class of constant-composition codes
have attracted recent interest due to its numerous applications,
such as in determining the zero error decision feedback capacity
of discrete memoryless channels \cite{b15}, multiple-access
communications \cite{b12}, spherical codes for modulation
\cite{b14}, DNA codes \cite{b11}, powerline communications
\cite{b13}, and frequency hopping \cite{b16}.

One of the most fundamental problem in coding theory is the
problem of determining the maximum size of a block code, given its
length and minimum distance. The problem of determining the
maximum size of a constant-composition code is much less
understood than the constant-weight and linear cases. In the
recent years, researches consider the problems of maximizing the
size of a constant-composition code (see \cite{b5, b2, b4}), and
constructing optimal codes to achieve these bounds (see \cite{b7,
b8, b10, b6, b9}). In this paper, we give a construction for
constant-composition codes then produce a lower bound on
constant-composition codes for arbitrary given minimum distance.
We show that when $q=3$ and $d=5$, our bound gives the best
possible size of constant-composition codes up to magnitude. As
far as we know, except for the bound given in this paper, there is
no bounds on $d>3$ so far.

This correspondence is organized as follows. In Section II, we
introduce some basic definitions and notations. We also review
some basic properties which will be used in this correspondence.
The main construction is presented in Section III. In Section IV,
Theorem 1 in section II are used to obtain some good lower bounds
on constant-composition codes.

\section{PRELIMINARY}

We use the standard notations for codes as follows. Let $\Z_{q}$
denote the set $\{0,1,\cdots,q-1\}$, and let $\Z_{q}^{n}$ be the
set of all $n$-tuples over $\Z_{q}$, where $q$ is a positive
integer. Let
$V_{n,[\omega_{0},\;\omega_{1},\;\cdots,\;\omega_{q-1}]}(q)$
denote the set of $n$-tuples over $\Z_{q}$ of the fixed
composition $[\omega_{0},\;\omega_{1},\;\cdots,\;\omega_{q-1}]$,
i.e., the number of $0$'s, $1$'s, $\cdots$, $q-1$'s in the
$n$-tuple over $\Z_{q}$ is given by
$\omega_{0},\omega_{1},\cdots,\omega_{q-1}$, respectively, where
$n=\omega_{0}+\omega_{1}+\cdots+\omega_{q-1}$. It is obvious that
$V_{n,[\omega_{0},\;\omega_{1},\;\cdots,\;\omega_{q-1}]}(q)$
contains $
\left(%
\begin{array}{c}
  n \\
  \omega_{0},\omega_{1},\cdots,\omega_{q-1} \\
\end{array}%
\right) $ elements. An
$(n,M,d,[\omega_{0},\;\omega_{1},\;\cdots,\;\omega_{q-1}])_{q}$
\textbf{constant-composition code} $C$ is a subset  of
$V_{n,[\omega_{0},\;\omega_{1},\;\cdots,\;\omega_{q-1}]}(q)$ with
size $M$ and minimum Hamming distance $d$. We use
$A_{q}(n,d,[\omega_{0},\;\omega_{1},\;\cdots,\;\omega_{q-1}])$ to
denote the maximum size of an
$(n,M,d,[\omega_{0},\;\omega_{1},\;\cdots,\;\omega_{q-1}])_{q}$
constant-composition code.\\ \indent In order to establish our
results in this correspondence, we need
the following Lemmas.\\
Let $\gcd(\alpha,\beta)$ be the greatest common divisor of the
positive integers $\alpha$ and $\beta$. Denote
\begin{equation}
Q=\prod_{{\footnotesize \begin{array}{c}\mbox{$p$ is prime}\\
    p\leq q-1\end{array}}}
 p , \;  {\rm for}
\;q\geq 3
\end{equation}
and
\begin{equation}
L(s,q)=\min\{l:\; l \geq s \;{\rm and} \;\gcd (l,Q)=1\}, {\rm
for}\; 0\leq s\leq Q-1\}.
\end{equation}
\begin{lemma}\label{lem1}(cf. \cite{b2})\\
\begin{equation}
A_{q}(n,3,[\omega_{0},\cdots,\omega_{q-1}])\geq
\left(%
\begin{array}{c}
  n\\
  \omega_{0},\cdots,\omega_{q-1}\\
\end{array}%
\right)/(n+\Gamma(t_{n},q)),
\end{equation}
where $t_{n} $ is the least nonnegative integer such that
$t_{n}\equiv n (\mod Q)$, and $\Gamma(t_{n},q)=L(t_{n},q)-t_{n}$.
\epf \end{lemma}
\begin{lemma}\label{lem2} (cf. \cite{b2}) Let $Q$ be given by (1).
If $\gcd(n,Q)=1$, then \begin{equation}
A_{q}(n,3,[\omega_{0},\cdots,\omega_{q-1}])\geq
\left(%
\begin{array}{c}
  n\\
  \omega_{0},\cdots,\omega_{q-1}\\
\end{array}%
\right)/n.
\end{equation}
\epf\end{lemma}
\begin{lemma}\label{lem3} (cf. \cite{b2}) For $q=3$,\\
$ A_{3}(n,3,[\omega_{0},\omega_{1},\omega_{2}])$\begin{equation}
\geq \left\{%
\begin{array}{ll}
   \left(%
\begin{array}{c}
  n\\
  \omega_{0},\omega_{1},\omega_{2}\\
\end{array}%
\right)/n, & \hbox{$n=2k+1$;} \\
    \left(%
\begin{array}{c}
  n\\
  \omega_{0},\cdots,\omega_{q-1}\\
\end{array}%
\right)/(n+1), & \hbox{$n=2k$.} \\
\end{array}%
\right.    \end{equation} \epf\end{lemma} In this correspondence,
bound (5) is improved for even length. \\
\indent For a constant-composition code with length $n$, minimum
distance at least $d$, and constant composition
$[\omega_{0},\cdots,\omega_{q-1}]$, denote
$\delta=\lfloor(d-1)/2\rfloor$. Then $\delta <
\omega_{1}+\cdots+\omega_{q-1}$.
\begin{lemma}\label{lem4} (cf. \cite{b2}) For any fixed $i$ where
$0 \leq i\leq q-1$, we have\\
$A_{q}(n,d,[\omega_{0},\cdots,\omega_{q-1}])$
\begin{equation}
\leq \left(%
\begin{array}{c}
  n\\
  \omega_{0},\cdots,\omega_{q-1}\\
\end{array}%
\right)/
\left(%
\begin{array}{c}
  \omega_{i}+\delta\\
  \omega_{i},\delta_{i,0},\cdots,\delta_{i,q-1}\\
\end{array}%
\right)
\end{equation}
where $\delta_{i,j},1\leq j\leq q-1$ are nonnegative integers such
that $\delta_{i,i} = 0$, $\delta_{i,0}+\cdots+\delta_{i,q-1}
=\delta$, and $\delta_{i,l}\leq \omega_{l}$ for $0\leq l\leq q-1$.
\epf\end{lemma} In this paper, we show that when $d=5$, we give a
lower bound have the same magnitude with bound (6).

\section{CONSTRUCTION OF CODES}
In this section, we generalize the construction that is proposed
by Xing \cite{b1}. Let $r$ be a prime power. We denote by
$\textbf{F}_{r}$ the finite field with $r$ elements. We label all
elements of $\textbf{F}_{r}$
$$\textbf{F}_{r}=\{\alpha_{0}=0,\alpha_{1},\cdots,\alpha_{r-1}\}.$$
For a positive integer $m$, consider the residue ring of
polynomials
$$\textbf{F}_{r}[x]/(x^{m}).$$
It is a finite ring and has $r^{m}$ elements. All invertible
elements of this ring form a multiplicative group, denoted by
$(\textbf{F}_{r}[x]/(x^{m}))^{\ast}$. It is a finite abelian
group. The quotient group
$$(\textbf{F}_{r}[x]/(x^{m}))^{\ast}/\textbf{F}_{r}^{\ast}$$
is a finite abelian group with $r^{m-1}$ elements. \\
Let $e$ is a positive integer, for a prime $p$, we define
$$\mu_{p}(e)=\left\{\begin{array}{cc}
  e & \;\text{if} \;p|e;\\
  e-1 & \;\text{otherwise}.\\
\end{array}\right.$$
\begin{theorem}\label{thm3.1} Let $q \geq 3$ be a integer and let $r$ be a power of $p$ for a prime $p$. If $p\geq q
$, then for any positive integer $d_{0}$ satisfying $1\leq
d_{0}\leq r-2$, there exist a $q$-ary $(r,M,\geq
\mu_{p}(d_{0})+2,[\omega_{0},\;\omega_{1},\;\cdots,\;\omega_{q-1}])$
constant-composition code
with\\
$$ M \geq \left(%
\begin{array}{c}
  r \\
  \omega_{0},\cdots,\omega_{q-1} \\
\end{array}%
\right)/r^{d_{0}-1}$$.
\end{theorem}
{\bf Proof.} Consider the map
$$\pi:V_{r,[\omega_{0},\;\omega_{1},\;\cdots,\;\omega_{q-1}]}(q)\rightarrow
(\textbf{F}_{r}[x]/x^{d_{0}})^{\ast}/\textbf{F}_{r}^{\ast}$$
$$(c_{1},c_{2},\cdots,c_{r})\mapsto
\overline{\prod_{i=1}^{r-1}(x-\alpha_{i})^{c_{i}}}.$$ By the
Pigeonhole Principle, it is clear that we can find one element
$\overline{f(x)}$ from this quotient
group such that it has at least $ \left(%
\begin{array}{c}
  r \\
  \omega_{0},\cdots,\omega_{q-1} \\
\end{array}%
\right)/r^{d_{0}-1}$
 pre-images, i.e., $\# (\pi^{-1}(\overline{f(x)}))
\geq
\left(%
\begin{array}{c}
  r \\
  \omega_{0},\cdots,\omega_{q-1} \\
\end{array}%
\right)/r^{d_{0}-1}$. Put
$$C=\pi^{-1}(\overline{f(x)}).$$
We are going to show that $C$ is a code with the desired
parameters. The length of $C$ is clearly $r$. The remaining thing
is to show that the minimum distance is at
least $\mu_{p}(d_{0})+2$.\\
\indent Let $\textbf{u}=(u_{1},u_{2},\cdots,u_{r})$ and
$\textbf{v}=(v_{1},v_{2},\cdots,v_{r})$ be two distinct codewords
of $C$. Then, $\pi(\textbf{u})=\pi(\textbf{v})=\overline{f(x)}.$
This implies that in the group
$(\textbf{F}_{r}[x]/(x^{d_{0}}))^\ast$, the element
$$\frac{\prod_{i=1}^{r-1}(x-\alpha_{i})^{u_{i}}}{\prod_{i=1}^{r-1}(x-\alpha_{i})^{v_{i}}}$$
is equal to $\alpha$ for some nonzero element $\alpha$ of
$\textbf{F}_{r}^{\ast}$.\\
Put
$$z:=\frac{\prod_{i=1}^{r-1}(x-\alpha_{i})^{u_{i}}}{\prod_{i=1}^{r-1}(x-\alpha_{i})^{v_{i}}}\in \textbf{F}_{r}(x).$$
It is clear that z is not a constant as $\textbf{u}\neq
\textbf{v}$.Then the principal divisor of $z$ is equal to \\
\begin{equation}
\textrm{div}(z)=\sum_{i=1}^{r-1}(u_{i}-v_{i})P_{i}+\left(\sum_{i=1}^{r-1}(v_{i}-u_{i})\right)P_{\infty}
\end{equation}
where $P_{i}$ is the place corresponding to $(x-\alpha_{i})$ for
all $1\leq i\leq r-1$, and $P_{\infty}$ is corresponding to the infinite place.\\
\indent Consider the field extension
$\textbf{F}_{r}(x)/\textbf{F}_{r}(z)$ of degree
\[\left(\sum_{i=1}^{r-1}|u_{i}-v_{i}|+\left |\sum_{i=1}^{r-1}(v_{i}-u_{i})\right |\right)/2\]
where $|\;.\;| $ stands for the absolute value of a real number.
We know
this extension is separable as $p\geq q$ (cf.\cite{b1}).\\
\indent For $1\leq i\leq r-1$, whenever $u_{i}-v_{i}\neq 0$, the
place $P_{i}$ has the ramification index $|u_{i}-v_{i}|$ in the
extension $\textbf{F}_{r}(x)/\textbf{F}_{r}(z)$ and hence the
different exponent $D_{P_{i}}$ of $P_{i}$ is at least
$|u_{i}-v_{i}|-1$ (see
\cite{b3}).\\
\indent The fact that $z$ is equal to $\alpha$ in the
group$(\textbf{F}_{r}[x]/(x^{d_{0}}))^\ast$ implies that $P_{0}$
is a zero of $z-\alpha$ with multiplicity at least $d_{0}$. Hence,
the ramification index of the place $P_{0}$ with respect to the
extension $\textbf{F}_{r}(x)/\textbf{F}_{r}(z)$ is at least
$d_{0}$, therefore, the different
exponent $D_{P_{0}}\geq d_{0}-1$. In particular, if $p|d_{0}$, by Dedekind's Different Theorem, we obtain $D_{P_{0}}\geq d_{0}$. So, $D_{P_{0}}\geq \mu_{p}(d_{0})$.\\
\indent Let
$$S=\{i\in\{1,2,\cdots,r-1\}:\;u_{i}\neq v_{i}\}$$ and let $\omega$ be the distance between $\textbf{u}$ and $\textbf{v}$.
\begin{itemize}\item[1]If $u_{r}=v_{r}$, then $|S|=\omega$ and
$\sum_{i=1}^{r-1}(v_{i}-u_{i})=0$. Hence
 $$\sum_{i\in S}D_{P_{i}}\geq
\sum_{i\in
S}(|u_{i}-v_{i}|-1)=\left(\sum_{i=1}^{r-1}|u_{i}-v_{i}|\right)-\omega.$$
By (7) the different exponent of $P_{\infty}$ with respect to the
extension $\textbf{F}_{r}(x)/\textbf{F}_{r}(z)$ at least 0. The
genera $g(\F_{r}(x))$ and $g(\F_{r}(z))$ are both equal to 0.
Thus, by the Huiwitz genus formula (see \cite{b3}), we have
{\small \begin{eqnarray*}
-2&=&2g(\F_{r}(x))-2\\
&=&(2g(\F_{r}(z))-2)[\F_{r}(x):\F_{r}(z)]+\sum_{P}D_{P}\\
&\ge &-2[\F_{r}(x):\F_{r}(z)]+\sum_{i\in S}D_{P_i}+D_{P_0}+D_{P_{\infty}}\\
 &\ge& -2
 \left(\left(\sum_{i=1}^{r-1}|u_{i}-v_{i}|\right)/2\right)+\left(\sum_{i=1}^{r-1}|u_i-v_i|\right)\\
 && -\omega+\mu_{p}(d_{0})\\
&=&\mu_{p}(d_{0})-\omega.
\end{eqnarray*}}
So, $\omega\geq \mu_{p}(d_{0})+2$. \item[2]If $u_{i}\neq v_{i}$,
then $|S|=\omega-1$. Hence
$$\sum_{i\in S}D_{P_{i}}\geq \sum_{i\in
S}(|u_{i}-v_{i}|-1)=\left(\sum_{i=1}^{r-1}|u_{i}-v_{i}|\right)-\omega+1.$$
By (7) the different exponent of $P_{\infty}$ with respect to the
extension $\textbf{F}_{r}(x)/\textbf{F}_{r}(z)$ at least
$\left|\sum_{i=1}^{r-1}(v_{i}-u_{i})\right|-1$. Thus by the
Huiwitz genus formula, we have {\small \begin{eqnarray*}
-2&=&2g(\F_{r}(x))-2\\
&=&(2g(\F_{r}(z))-2)[\F_{r}(x):\F_{r}(z)]+\sum_{P}D_{P}\\
&\ge &-2[\F_{r}(x):\F_{r}(z)]+\sum_{i\in S}D_{P_i}+D_{P_0}+D_{P_{\infty}}\\
 &\ge&-2\left(\left(\sum_{i=1}^{r-1}|u_{i}-v_{i}|+\left |\sum_{i=1}^{r-1}(v_{i}-u_{i})\right |\right)/2\right)+\\
&&\left(\sum_{i=1}^{r-1}|u_i-v_i|\right)-\omega+\mu_{p}(d_{0})+\left|\sum_{i=1}^{r-1}(v_{i}-u_{i})\right|-1\\
&=&\mu_{p}(d_{0})-\omega.
\end{eqnarray*}}
So, $\omega\geq \mu_{p}(d_{0})+2$.
\end{itemize}
The desired result follows.\epf

\section{SOME EXAMPLES FOR LOWER BOUND ON CONSTANT-COMPOSITION CODES}
Now, we can get some improved lower bounds for
constant-composition codes from Theorem 1. We adopt the notations
and terminologies in the previous section and consider the
quotient group
$$(\textbf{F}_{r}[x]/x^{d_{0}})^{\ast}/\textbf{F}_{r}^{\ast}.$$
\textbf{Example 1}.\; Consider $d_{0}=2$ \begin{itemize}
\item[(1)] For the case $p\geq q\geq 3$, $\mu_{p}(d_{0})= 1$, the
group $(\F_{r}[x]/(x^{2}))^{\ast}/(\F_{r})^{\ast}$ has $r$
elements. By Theorem 1  we can get a constant-composition code
with parameters
$(r,M,d,[\omega_{0},\;\omega_{1},\;\cdots,\;\omega_{q-1}])$, where
$d\geq 3$, and
$$M \geq\left(
\begin{array}{c}
  r \\
  \omega_{0},\cdots,\omega_{q-1} \\
\end{array}
\right)/r. $$ Then we obtain
\begin{equation}
A_{q}(r,3,[\omega_{0},\cdots,\omega_{q-1}])\geq \left(
\begin{array}{c}
  r \\
  \omega_{0},\cdots,\omega_{q-1} \\
\end{array}
\right)/r.\end{equation}
The bound in this case achieves the one
given in Lemma 2 for codes with odd length.
\item[(2)] Now we
consider the code of even length. Let $q=3$, $2|r$, from the first
part proof of theorem 1, we know that we can get a
constant-composition code of size $\geq \left(
\begin{array}{c}
  r \\
  \omega_{0},\omega_{1},\omega_{2} \\
\end{array}
\right)/r$, then we want to show this code has minimum distance
$\geq 3$. For two distinct codewords
$\textbf{u}=(u_{1},u_{2},\cdots,u_{r})$ and
$\textbf{v}=(v_{1},v_{2},\cdots,v_{r})$, similar to Theorem 1,
consider\begin{equation}\frac{u(x)}{v(x)}:=\frac{\prod_{i=1}^{r-1}(x-\alpha_{i})^{u_{i}}}{\prod_{i=1}^{r-1}(x-\alpha_{i})^{v_{i}}}\equiv
\alpha \;{\rm mod}(x^{2}).\end{equation} for some nonzero element
$\alpha$ of $\F_{r}^{\ast}$.
\begin{itemize}\item[1]If $u_{r}\neq v_{r}$, the distance between $\textbf{u}$ and $\textbf{v}$ is 2 if and only if $\frac{u(x)}{v(x)}=(x-\alpha_{i})$ or $\frac{u(x)}{v(x)}=\frac{1}{x-\alpha_{i}}$ for some $i,\;1\leq i\leq r-1.$ Both of these two cases are not satisfy (9), so we get $d\geq 3$.
\item[2]If $u_{r}=v_{r}$, it is easy to know that the distance
between $\textbf{u}=(u_{1},u_{2},\cdots,u_{r})$ and
$\textbf{v}=(v_{1},v_{2},\cdots,v_{r})$ is $2$ if and only if
$\frac{u(x)}{v(x)}=\frac{x-\alpha_{i}}{x-\alpha_{j}}$ or
$\frac{u(x)}{v(x)}=\frac{(x-\alpha_{i})^{2}}{(x-\alpha_{j})^{2}}$,
for some $i,j,1\leq i\neq j\leq r-1$. Since ${\rm
char}{\F_{r}}=2$, both of these two cases are not satisfy $(9)$,
so we get $d\geq 3$.\end{itemize} Then\begin{equation}
A_{3}(r,3,[\omega_{0},\omega_{1},\omega_{2}])\geq \left(
\begin{array}{c}
  r \\
  \omega_{0},\omega_{1},\omega_{2} \\
\end{array}
\right)/r.\end{equation} Bound (10) improves the one given in
Lemma $3$ when the length of code is even.
\end{itemize}
\bigskip
 \noindent\textbf{Example 2}. Consider $d_{0}=3$:
\begin{itemize}\item[(1)]\; For the case $p=\textrm{char}(\F_{r})=q=3$. Then
$p|d_{0}$, since $\mu_{p}(d_{0})=3$ we get $d\geq 5$. By Theorem
1, we get a $3$-ary $(r,M,5,[\omega_{0},\omega_{1},\omega_{2}])$
constant-composition code, where
$$ M\geq\left(
\begin{array}{c}
  r\\
  \omega_{0},\omega_{1},\omega_{2} \\
\end{array}
\right)/r^{2}.$$  Hence,
\begin{equation}
A_{3}(r,5,[\omega_{0},\omega_{1},\omega_{2}])\geq \left(
\begin{array}{c}
  r\\
  \omega_{0},\omega_{1},\omega_{2} \\
\end{array}
\right)/r^{2}.\end{equation}  Lemma 3 given a upper bound of
constant-composition codes. Now we take $d=5$, then
$\delta=\lfloor\frac{d-1}{2}\rfloor=2$, it is easy to know that
there exist $\omega_{i}\geq \lfloor r/q\rfloor$ for $0\leq i\leq
q-1$. So we have
$$A_{3}(r,5,[\omega_{0},\omega_{1},\omega_{2}])\leq \left(%
\begin{array}{c}
  n\\
  \omega_{0},\omega_{1},\omega_{2}\\
\end{array}%
\right)/
\left(%
\begin{array}{c}
  \omega_{i}+2\\
  \omega_{i},\delta_{i,0},\delta_{i,1},\delta_{i,2}\\
\end{array}%
\right)$$ where $\delta_{i,j}$ are nonnegative $\delta_{i,0}+\delta_{i,1}+\delta_{i,2}=2$, we choose $\delta_{i,0},\delta_{i,1},\delta_{i,2}$ such that $\left(%
\begin{array}{c}
  2\\
  \delta_{i,0},\delta_{i,1},\delta_{i,2}\\
\end{array}%
\right)=2$, then $t(r)=\left(%
\begin{array}{c}
  \omega_{i}+2\\
  \omega_{i},\delta_{i,0},\delta_{i,1},\delta_{i,2}\\
\end{array}%
\right)= (\omega_{i}+2)(\omega_{i}+1)\geq
(\frac{r}{q}+1)\frac{r}{q}= O(r^{2})$ when $r\rightarrow \infty$,
then we obtain an upper bound for constant composition code over
$\F_{3}$ of minimum distance $5$\\
$$A_{3}(r,5,[\omega_{0},\omega_{1},\omega_{2}])\leq  \left(%
\begin{array}{c}
  r\\
  \omega_{0},\omega_{1},\omega_{2}\\
\end{array}%
\right)/t(r)$$ where $t(r)=O(r^{2})$, compare this upper bound
with our lower bound in (9), our lower bound given the best
possible size up to magnitude. \item[(2)]\; For the case $p\geq
q\geq 3$ and $p>3$, then we obtain $d\geq 4$ since
$\mu_{p}(d_{0})=2$, By Theorem 1, we obtain a $q$-ary
$(r,M,4,[\omega_{0},\;\omega_{1},\;\cdots,\;\omega_{q-1}])$ constant-composition code, where \\
$$M \geq\left(
\begin{array}{c}
  r\\
  \omega_{0},\cdots,\omega_{q-1} \\
\end{array}
\right)/r^{2},$$ and the lower bound
\begin{equation}
A_{q}(r,4,[\omega_{0},\cdots,\omega_{q-1}])\geq \left(
\begin{array}{c}
  r\\
  \omega_{0},\cdots,\omega_{q-1} \\
\end{array}
\right)/r^{2}.\end{equation} \end{itemize}
\bigskip
 \noindent \textbf{Example 3}.\; Let
 $d_{0}=5$\begin{itemize}\item[(1)]
For the case $p= q=5$, we get $d\geq 7$, then by Theorem 1,\\
$$A_{q}(r,7,[\omega_{0},\cdots,\omega_{q-1}])\geq\left(
\begin{array}{c}
  r \\
  \omega_{0},\cdots,\omega_{q-1} \\
\end{array}
\right)/r^{4}.$$
 \item[(2)]If $p=>5$ and $3\leq q\leq p$, we get $d\geq
 6$ and lower bound
$$A_{q}(r,6,[\omega_{0},\cdots,\omega_{q-1}])\geq\left(
\begin{array}{c}
  r \\
  \omega_{0},\cdots,\omega_{q-1} \\
\end{array}
\right)/r^{4}.$$
\end{itemize}
\bigskip

\textbf{Remark 2:} \begin{itemize} \item[1]The construction in
this paper produces a lower bound on constant-composition codes
for arbitrary given minimum distance. \item[2] As far as we know,
except for the bound given in this paper, there are no bounds on
$A_{q}(n,d,[\omega_{0},\cdots,\omega_{q-1}])$, where $d\geq4$, so
far.
\end{itemize}

\section*{Acknowledgment}
The author is grateful to Profs. Keqin Feng, Jianlong Chen and
Chaoping Xing for their guidance.

\end{document}